# Enriched physics-informed neural networks for in-plane crack problems: Theory and MATLAB codes

Yan Gu[1,2], Chuanzeng Zhang[3*], Peijun Zhang[3], Mikhail V. Golub[4]

[1]*School of Mathematics and Statistics, Qingdao University, Qingdao 266071, PR China*

[2]*Institute of Mechanics for Multifunctional Materials and Structures, Qingdao University, Qingdao 266071, PR China*

[3]*Department of Civil Engineering, University of Siegen, Paul-Bonatz-Str. 9-11, D-57076 Siegen, Germany*

[4]*Institute for Mathematics, Mechanics and Informatics, Kuban State University, Krasnodar, 350040, Russian Federation*

## Abstract

In this paper, a method based on the physics-informed neural networks (PINNs) is presented to model in-plane crack problems in the linear elastic fracture mechanics. Instead of forming a mesh, the PINNs is meshless and can be trained on batches of randomly sampled collocation points. In order to capture the theoretical singular behavior of the near-tip stress and strain fields, the standard PINNs formulation is enriched here by including the crack-tip asymptotic functions such that the singular solutions at the crack-tip region can be modeled accurately without a high degree of nodal refinement. The learnable parameters of the enriched PINNs are trained to satisfy the governing equations of the cracked body and the corresponding boundary conditions. It was found that the incorporation of the crack-tip enrichment functions in PINNs is substantially simpler and more trouble-free than in the finite element (FEM) or boundary element (BEM) methods. The present algorithm is tested on a class of representative benchmarks with different modes of loading types. Results show that the present method allows the calculation of accurate stress intensity factors (SIFs) with far fewer degrees of freedom. A self-contained MATLAB code and data-sets accompanying this manuscript are also provided.

* Corresponding Author: c.zhang@uni-siegen.de

# 1. Introduction

The fracture mechanics analysis is one of the most cumbersome tasks in the community of computational mechanics. The fractured domain causes a singularity in the stress field where the stress may reach an infinite value at the crack-tip region. Currently, extended/generalized finite element methods (FEM) based on special crack-tip elements (for example the quarter-point element, etc.) as well as the enriched elements are widely used in elastic fracture analysis [1-4]. The method based on the special crack-tip elements is to replace the original shape functions in the polynomial space by special shape functions that can capture the theoretical singular behavior of the near-tip solutions. Enriched element methods, which incorporate the asymptotic crack-tip fields in the trial functions, are able to provide accurate stress intensity factors (SIFs) directly as part of the solution. Both of these methods can achieve high accurate SIFs results without a high degree of near-tip mesh refinement. In spite of the great successes using finite elements, mesh generation in three dimensions is still time consuming and especially burdensome for crack growth simulations.

The boundary element method (BEM) can be viewed as an important alternative to the FEM for fracture mechanics analysis due to its boundary-only discretization and semi-analytical nature. The modeling of crack problems using the BEM with remeshing and/or various enriched techniques has been pursued by many researchers [5-12]. However, the requirement of the fundamental solution (Green's function) imposes restriction on the scope of application of the BEM-based methods. These methods are also not readily extendable to non-linear problems nor to the study of cracks in general anisotropic materials. More recently, some of the meshless methods, such as the element-free Galerkin (EFG) method [2, 13] and the generalized finite difference method (GFDM) [14], have been successful in modelling static and dynamic fracture problems in two- and three-dimensions.

In this work we propose to solve in-plane crack problems in linear elastic fracture mechanics by using the physics-informed neural networks (PINNs). The methods based on the PINNs have recently received great attention thanks to their flexibility in solving a wide range of partial differential equations (PDEs) [15-18]. The method seamlessly integrates the information from both the physical laws and the measurement data by embedding the underlying PDEs and the corresponding boundary/initial conditions into the loss functions of a neural network. The unknown parameters of the neural network can be well-trained by minimizing the loss functions via various gradient-based optimization techniques. Some attractive features of the PINNs are summarized as follows: (1) Compared with the traditional numerical methods, the methods based on the PINNs are meshless and thus can easily handle complex and moving-domain problems; (2) The employment of PINNs provides closed-form and differentiable solutions which are easily used in any subsequent calculation; (3) By enforcing/embedding the physical laws (the underlying PDEs), the PINNs can be trained with a small amount of measurement data; (4) The code is easier to implement by using existing open-source deep learning frameworks.

In the standard PINNs formulation with conventional neural networks, we found that the PINNs solution may initially converge by increasing the number of the neural units, but eventually diverges from the theoretical solution. This is because of the fact that the standard neural networks cannot basically capture the theoretical singular behavior predicted for the stress and strain fields, even with a high degree of mesh refinement at the crack-tip region. In order to overcome this deficiency, we present in this study an enriched PINNs formulation by incorporating the crack-tip asymptotic/enrichment functions such that the displacement and stress distributions at the crack-tip region can be modeled accurately without extremely refined meshes. One of the advantages of this method is that the SIFs can be computed directly as part of the learnable parameters. It was also found that the incorporation of the crack-tip enrichment functions in the PINNs is substantially simpler and more trouble-free than in the FEM or BEM. The numerical results presented in this paper demonstrate that for in-plane crack problems, the accuracy and performance of the enriched PINNs are excellent which open up many interesting possibilities for its further development and applications.

The outline of this paper is as follows. The governing equations of a cracked body and the near-tip displacement and stress fields are briefly introduced in Section2. Section 3 presents a brief overview of the standard PINNs for general 2D boundary value problems, and in Section 4, the computational methodology adopted in the enriched PINNs and the numerical issues involved are discussed. Next in Section 5, we perform four representative benchmarks to verify the correctness of the present method. The numerical SIF results are also compared to available BEM reference solutions from the literature. Some final conclusions and remarks are provided in Section 6. Finally, a self-contained MATLAB code accompanying this manuscript are provided in Appendix A.

## 2. Problem statements

### 2.1. Governing equations of a cracked body

For a homogeneous, isotropic and linear elastic solid, the strong form of the equilibrium equations of the cracked body can be expressed as:

$$\frac{\partial \sigma_{11}}{\partial x_1} + \frac{\partial \sigma_{12}}{\partial x_2} + f_1 = 0, \tag{1}$$

$$\frac{\partial \sigma_{21}}{\partial x_1} + \frac{\partial \sigma_{22}}{\partial x_2} + f_2 = 0, \tag{2}$$

or in tensor notation:

$$\sigma_{ij,j} + f_i = 0, \quad i,j = 1,2, \tag{3}$$

where $\sigma_{11}$ and $\sigma_{22}$ denote the direct stresses in the $x_1$ and $x_2$ directions, respectively, $\sigma_{12}$ or $\sigma_{21}$ stands for the shear stresses, $f_1$ and $f_2$ are the $x_1$ and $x_2$ components of the body force loading applied to the structure. The stress-strain relationships (Hooke's law) for plane strain assumption can be written as follows (in tensor notation):

$$\sigma_{ij} = 2\mu \left( \varepsilon_{ij} + \frac{v}{1-2v} \varepsilon_{kk} \delta_{ij} \right), \quad i,j = 1,2, \tag{4}$$

where $v$ is Poisson's ratio, $\mu$ is the shear modulus defined as $\mu = \dfrac{E}{2(1+v)}$ in which $E$ denotes the Young's modulus, $\delta_{ij}$ is the Kronecker delta, and the strains $\varepsilon_{ij}$ are related to displacements according to the following strain-displacement relationships:

$$\varepsilon_{ij} = \frac{1}{2}\left( u_{i,j} + u_{j,i} \right), \tag{5}$$

where $u_1$ and $u_2$ are the $x_1$ and $x_2$ components of the displacement vector. The equilibrium equations and can be expressed in terms of displacement components as follows:

$$\mu \frac{\partial^2 u_1}{\partial x_1^2} + \mu \frac{\partial^2 u_1}{\partial x_2^2} + \frac{\mu}{1-2v}\left( \frac{\partial^2 u_1}{\partial x_1^2} + \frac{\partial^2 u_2}{\partial x_1 \partial x_2} \right) = -f_1, \tag{6}$$

$$\mu \frac{\partial^2 u_2}{\partial x_1^2} + \mu \frac{\partial^2 u_2}{\partial x_2^2} + \frac{\mu}{1-2v}\left( \frac{\partial^2 u_2}{\partial x_2^2} + \frac{\partial^2 u_1}{\partial x_1 \partial x_2} \right) = -f_2. \tag{7}$$

The above equations have to be considered together with the following displacement and/or traction boundary conditions:

$$u_i = \overline{u}_i, \quad (Dirichlet), \tag{8}$$

$$t_i = \sigma_{ij} n_j = \overline{t}_i, \quad (Neumann), \tag{9}$$

where $n_i$ and $t_i$ denote the unit outward normal and the traction component in the $x_i$ direction,

respectively, $\bar{u}_i$ and $\bar{t}_i$ stand for the prescribed displacement and traction at the boundary.

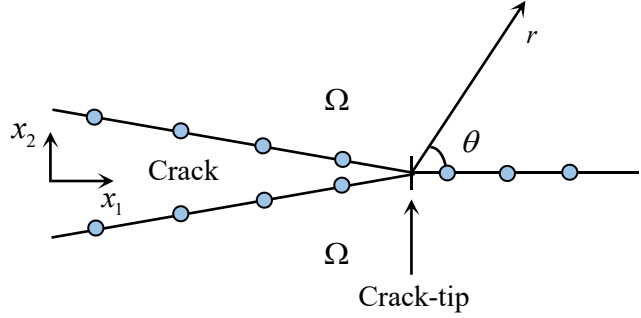

**Fig. 1.** The crack-tip region of a fractured domain.

## 2.2. The stress intensity factors and the near-tip displacement and stress fields

In the linear elastic fracture mechanics, the displacement, stress and strain fields can be determined by employing the concept of the SIFs near the crack-tip region. It is therefore important to accurately calculate the SIFs for fracture mechanics analysis. Referring to Fig. 1 for notation, the displacement and stress fields in the vicinity of the crack-tip can be expressed as (with higher-order terms omitted) [19-21]:

$$\begin{Bmatrix} u_1 \\ u_2 \end{Bmatrix} = \frac{K_I}{2\mu}\sqrt{\frac{r}{2\pi}}(\kappa - \cos\theta)\begin{Bmatrix} \cos(\theta/2) \\ \sin(\theta/2) \end{Bmatrix}, \tag{10}$$

$$\begin{Bmatrix} \sigma_{11} \\ \sigma_{22} \\ \sigma_{12} \end{Bmatrix} = \frac{K_I}{\sqrt{2\pi r}}\cos\left(\frac{\theta}{2}\right)\begin{Bmatrix} 1-\sin(\theta/2)\sin(3\theta/2) \\ 1+\sin(\theta/2)\sin(3\theta/2) \\ \sin(\theta/2)\cos(3\theta/2) \end{Bmatrix}, \tag{11}$$

for mode I crack problem and

$$\begin{Bmatrix} u_1 \\ u_2 \end{Bmatrix} = \frac{K_{II}}{2\mu}\sqrt{\frac{r}{2\pi}}\begin{Bmatrix} \sin(\theta/2)(\kappa+2+\cos\theta) \\ \cos(\theta/2)(\kappa-2+\cos\theta) \end{Bmatrix}, \tag{12}$$

$$\begin{Bmatrix} \sigma_{11} \\ \sigma_{22} \\ \sigma_{12} \end{Bmatrix} = \frac{K_{II}}{\sqrt{2\pi r}}\begin{Bmatrix} -\sin(\theta/2)[2+\cos(\theta/2)\cos(3\theta/2)] \\ \sin(\theta/2)\cos(\theta/2)\cos(3\theta/2) \\ \cos(\theta/2)[1-\sin(\theta/2)\sin(3\theta/2)] \end{Bmatrix}, \tag{13}$$

for mode II crack problem, where $(r,\theta)$ stands for the polar coordinate system of the calculated node about the crack-tip, and

$$\kappa = \begin{cases} 3-4\nu, & \text{for plane strain}, \\ (3-\nu)/(1+\nu), & \text{for plane stress}. \end{cases} \quad (14)$$

In the above relations -, the parameters $K_I$ and $K_{II}$ are SIFs in modes I and II, respectively, defined as:

$$K_I = \lim_{\substack{r\to 0 \\ \theta=0}} \sigma_{22}\sqrt{2\pi r}, \quad K_{II} = \lim_{\substack{r\to 0 \\ \theta=0}} \sigma_{12}\sqrt{2\pi r}, \quad (15)$$

The SIFs represent the strength of the displacement and stress fields surrounding the crack-tip. It can be seen from relations and that the stresses $\sigma_{11}$, $\sigma_{22}$ and $\sigma_{12}$ are singular near the crack-tip when $r \to 0$. The modes I and II SIFs $K_I$ and $K_{II}$ can be finally calculated by evaluating the displacement and stress fields at the crack-tip region, and by substituting these near-tip variables into relations -.

## 3. The basis of the standard physics-informed neural networks (PINNs)

To understand the key ideas clearly, let's considered the following 2D boundary value problem of the general form:

$$Lu(\boldsymbol{x}) = f(\boldsymbol{x}), \quad (16)$$

$$u(\boldsymbol{x}) = \bar{u}, \quad (Dirichlet), \quad (17)$$

$$\frac{\partial u(\boldsymbol{x})}{\partial \boldsymbol{n}} = \bar{q}, \quad (Neumann), \quad (18)$$

where $L$ denotes a differential operator, $u(\boldsymbol{x})$, $\boldsymbol{x}=(x_1,x_2)\in\Omega$, is the solution to be calculated. We remark here that for dynamic problems, we can consider the time variable $t$ as an additional coordinate in $\boldsymbol{x}$, i.e., $(x_1,x_2,t)$, and the computational domain $\Omega$ becomes the spatio-temporal domain. Here we propose to solve the above-mentioned boundary value problem by using the PINNs. The method seamlessly integrates the information from both the physical laws and the measurement data by embedding the underlying PDEs and the corresponding boundary conditions into the loss functions of a neural network. For simplicity, let us start by considering a simple neural architecture with three fully connected operations, i.e., the input channels corresponding to the inputs $x_1$ and $x_2$, one hidden layer with $M$ neuron units, and one output channel corresponding to the target solution $u(x)$.

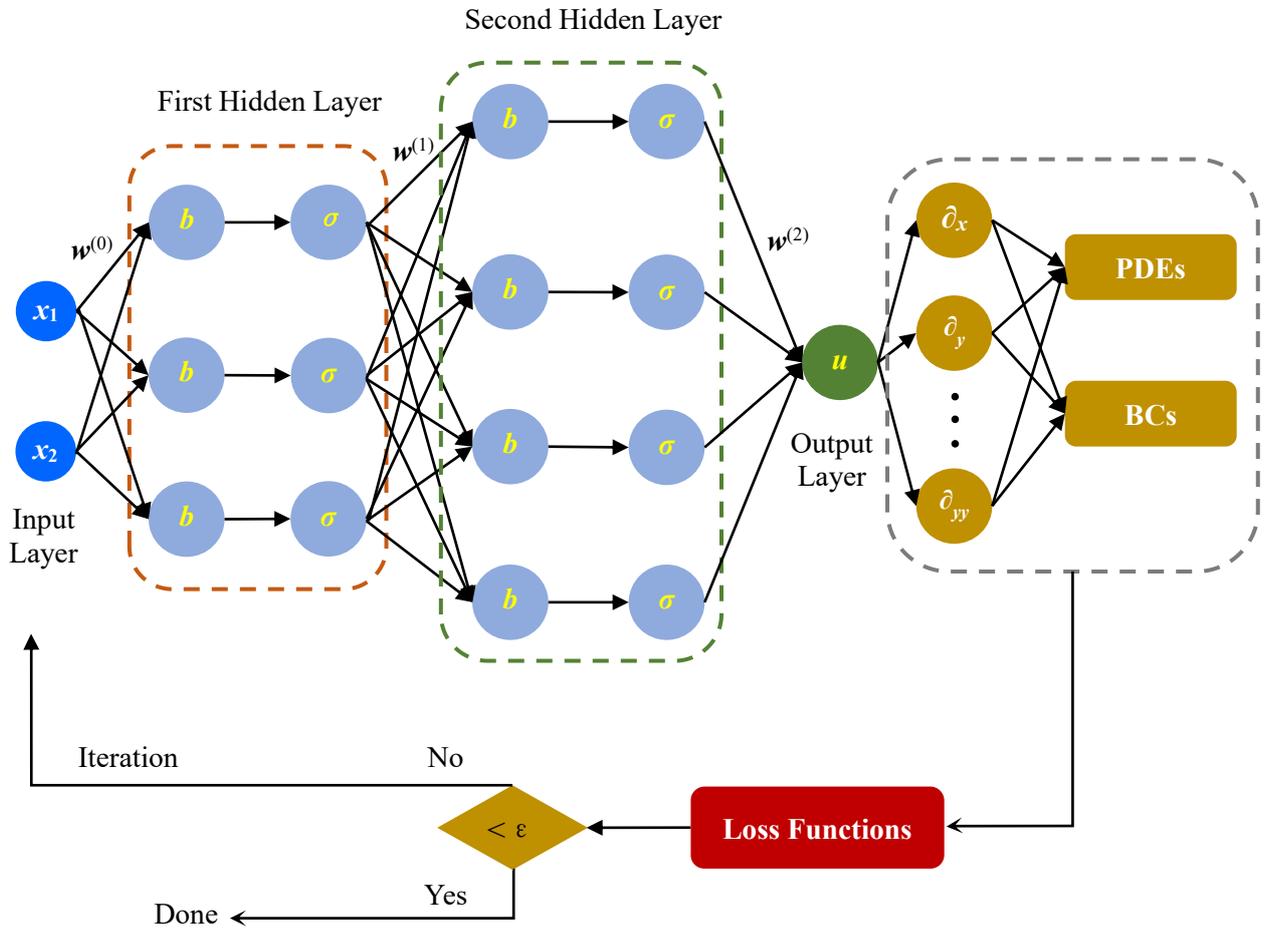

**Fig. 2.** The framework of the physics-informed neural networks (PINNs).

Following the pioneering works by Lagaris et al. [15] and Raissi et al. [16], the target solution $u(x)$ can be approximated by using the following network with a set of trainable parameters (i.e., the weights $w$ and biases $b$ of the neural architecture):

$$u(\boldsymbol{x},\boldsymbol{w},\boldsymbol{b}) = \sum_{i=1}^{M} w_i^{(1)} \sigma\left(z_i^{(0)}\right) + b^{(1)}, \tag{19}$$

$$z_i^{(0)} = \sum_{j=1}^{2} w_{ij}^{(0)} x_j + b_i^{(0)} = w_{i1}^{(0)} x_1 + w_{i2}^{(0)} x_2 + b_i^{(0)}, \tag{20}$$

where $w_{ij}^{(0)}$ denotes the weight from the input unit $x_j$ to the $i^{th}$ neuron unit, $w_i^{(1)}$ stands for the weight from the $i^{th}$ neuron unit to the output, i.e., $u(\boldsymbol{x},\boldsymbol{w},\boldsymbol{b})$, $b_i^{(0)}$ and $b^{(1)}$ are the biases of the $i^{th}$ neuron unit and the output layer, respectively, and $\sigma(\cdot)$ denotes the nonlinear activation function [22]. The extension to the case of more hidden-layers can be obtained accordingly [22, 23]. Fig. 2 gives a sketch of a PINNs architecture with two hidden-layers.

In the PINNs, the trainable parameters $\boldsymbol{w}$ and $\boldsymbol{b}$ in the network can be "learned" by substituting the relations    and    into the governing equation and the corresponding boundary conditions, and by minimizing the following loss function:

$$L(\boldsymbol{w},\boldsymbol{b}) = L_{PDE}(\boldsymbol{w},\boldsymbol{b}) + \lambda_u L_u(\boldsymbol{w},\boldsymbol{b}) + \lambda_q L_q(\boldsymbol{w},\boldsymbol{b}), \tag{21}$$

in which

$$L_{PDE}(\boldsymbol{w},\boldsymbol{b}) = \frac{1}{N_{PDE}} \sum_{i=1}^{N_{PDE}} \left| Lu(\boldsymbol{x}_{PDE}^i, \boldsymbol{w}, \boldsymbol{b}) - f(\boldsymbol{x}_{PDE}^i) \right|^2, \tag{22}$$

$$L_u(\boldsymbol{w},\boldsymbol{b}) = \frac{1}{N_u} \sum_{i=1}^{N_u} \left| u(\boldsymbol{x}_u^i, \boldsymbol{w}, \boldsymbol{b}) - \bar{u}_i \right|^2, \tag{23}$$

$$L_q(\boldsymbol{w},\boldsymbol{b}) = \frac{1}{N_q} \sum_{i=1}^{N_q} \left| \frac{\partial u(\boldsymbol{x}_q^i, \boldsymbol{w}, \boldsymbol{b})}{\partial \boldsymbol{n}} - \bar{q}_i \right|^2, \tag{24}$$

where the mean squared errors $L_{PDE}(\boldsymbol{w},\boldsymbol{b})$, $L_u(\boldsymbol{w},\boldsymbol{b})$ and $L_q(\boldsymbol{w},\boldsymbol{b})$ correspond to the losses of the underlying PDE, Dirichlet and Neumann boundary conditions, respectively, $\boldsymbol{x}_{PDE}$, $\boldsymbol{x}_u$ and $\boldsymbol{x}_q$ are three sets of training points sampled in the interior of the domain and on the Dirichlet and Neumann boundary locations, respectively, $N_{PDE}$, $N_u$ and $N_q$ stand for the batch sizes for the training data $\boldsymbol{x}_{PDE}$, $\boldsymbol{x}_u$ and $\boldsymbol{x}_q$, respectively, $\lambda_u$ and $\lambda_q$ are two hyper-parameters used to balance the influence between different loss terms during the training. The hyper-parameters $\lambda_u$ and $\lambda_q$, which can be either user-defined or tuned automatically during the training, have a crucial impact on improving the performance of the PINNs. It is noted here that the optimal choice of these parameters is still an open problem in the PINNs literature and we refer the interested reader to Refs. [17, 23, 24] for more details. In our computations, we choose $\lambda_u = \lambda_q = 1$. The loss function measures how well the trial solution $u(\boldsymbol{x},\boldsymbol{w},\boldsymbol{b})$ satisfies the underlying PDE and the corresponding boundary conditions. According to the above analysis, the original PDE problem can now be recast into an optimization problem, i.e., the network is trained by minimizing the loss function via various gradient-based optimization techniques.

## 4. The enriched PINNs for in-plane crack problems

The governing equations of the cracked body studied in this paper involve the coupling of two PDEs. One way to treat this problem is to define two different neural networks for each physics, i.e., two displacement components, separately and then coupled them together in a single loss function. This kind of model is accurate but the training process could be computationally expensive. Another way is to use only a single neural network for all of the physics such that the outputs of the PINNs can be approximated by using the same set of trainable parameters (see Fig. 3). We here focus on the second strategy and the first one will not be discussed. In the standard PINNs formulation, the displacement components $u_1$ and $u_2$ can be approximated by using the following neural networks:

$$u_1(\boldsymbol{x},\boldsymbol{w},\boldsymbol{b}) = \sum_{i=1}^{M} w_{1i}^{(1)} \sigma\left(z_i^{(0)}\right) + b^{(1)}, \tag{25}$$

$$u_2(\boldsymbol{x},\boldsymbol{w},\boldsymbol{b}) = \sum_{i=1}^{M} w_{2i}^{(1)} \sigma\left(z_i^{(0)}\right) + b^{(2)}, \tag{26}$$

where again $z_i^{(0)} = w_{i1}^{(0)}x_1 + w_{i2}^{(0)}x_2 + b_i^{(0)}$, $M$ is the number of neuron units employed in the hidden layer, $\{w_{ij}^{(0)}\}_{j=1,2}$ denote the weights from the input unit $x_j$ to the $i^{\text{th}}$ neuron unit, $\{w_{ji}^{(1)}\}_{j=1,2}$ stand for the weight from the $i^{\text{th}}$ neuron unit to the output $u_j$, $b_i^{(0)}$ stands for the bias of the $i^{\text{th}}$ neuron unit, and $b^{(1)}$ and $b^{(2)}$ are the biases corresponding to the outputs $u_1$ and $u_2$, respectively.

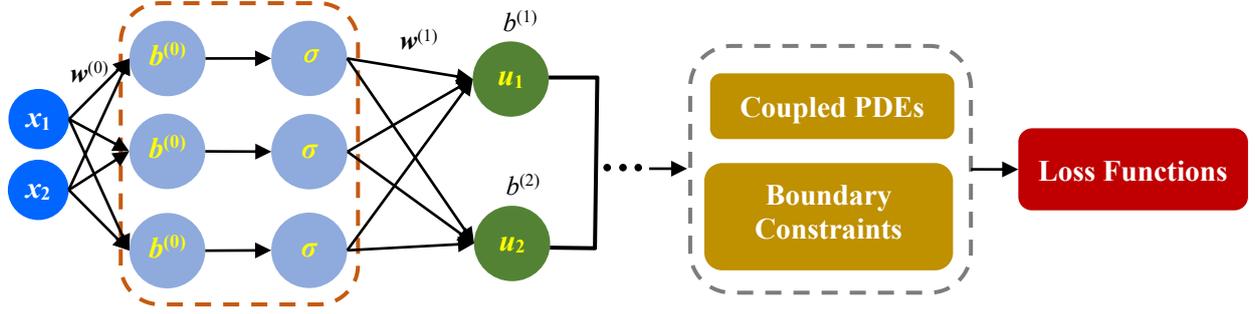

**Fig. 3.** The process of the PINNs model for 2D elasticity problems.

As indicated above, modeling crack problems by using the standard PINNs models is inadequate and may lead to inaccurate results, even a high degree of mesh refinement is employed. This is because of the fact that the standard neural networks cannot basically capture the theoretical singular behavior predicted for the stress and strain fields. In order to overcome this deficiency, special functions, which depend on the nature of the near-tip behavior, must be included into the PINNs approximation, such that the crack-tip fields can be modeled accurately without extremely refined meshes. With this idea in mind, the enriched displacement field can be rewritten by employing the crack-tip asymptotic/enrichment functions as:

$$u_1(\boldsymbol{x}, \boldsymbol{w}, \boldsymbol{b}, \tilde{K}_{\text{I}}, \tilde{K}_{\text{II}}) = \left[\sum_{i=1}^{M} w_{1i}^{(1)} \sigma(z_i^{(0)}) + b^{(1)}\right] \\ + \tilde{K}_{\text{I}}\left[\sqrt{r}(\kappa - \cos\theta)\cos(\theta/2)\right] \\ + \tilde{K}_{\text{II}}\left[\sqrt{r}\sin(\theta/2)(\kappa + 2 + \cos\theta)\right], \quad (27)$$

$$u_2(\pmb{x},\pmb{w},\pmb{b},\tilde{K}_\mathrm{I},\tilde{K}_\mathrm{II}) = \left[\sum_{i=1}^{M} w_{2i}^{(1)}\sigma\left(z_i^{(0)}\right)+b^{(2)}\right]$$
$$+ \tilde{K}_\mathrm{I}\left[\sqrt{r}(\kappa-\cos\theta)\sin(\theta/2)\right] \qquad (28)$$
$$+ \tilde{K}_\mathrm{II}\left[\sqrt{r}\cos(\theta/2)(\kappa-2+\cos\theta)\right],$$

where the second and third terms on the right-hand side of the above equations are the enriched functions, which can be found in Section 2.2, $\tilde{K}_\mathrm{I}$ and $\tilde{K}_\mathrm{II}$ are two hyper-parameters which can be learned automatically during the training process, $r$ and $\theta$ are polar coordinates of the calculation point $\pmb{x}$ about the crack-tip. Accordingly, the stress and strain vectors corresponding to the enriched displacement field can be obtained by differentiating the enriched neural networks with respect to their input coordinates. In our computations, the automatic differentiation function "dlgradient" is used to differentiate the neural networks (requires the Deep Learning Toolbox in the MATLAB). On substituting the enriched displacement functions and their derivatives into the governing equations as well as the corresponding boundary conditions, the following loss function can be obtained:

$$L(\pmb{w},\pmb{b},\tilde{K}_\mathrm{I},\tilde{K}_\mathrm{II}) = L_{PDE}^{(1)}(\pmb{w},\pmb{b},\tilde{K}_\mathrm{I},\tilde{K}_\mathrm{II}) + L_{PDE}^{(2)}(\pmb{w},\pmb{b},\tilde{K}_\mathrm{I},\tilde{K}_\mathrm{II})$$
$$+ L_{u_1}(\pmb{w},\pmb{b},\tilde{K}_\mathrm{I},\tilde{K}_\mathrm{II}) + L_{u_2}(\pmb{w},\pmb{b},\tilde{K}_\mathrm{I},\tilde{K}_\mathrm{II}) \qquad (29)$$
$$+ L_{t_1}(\pmb{w},\pmb{b},\tilde{K}_\mathrm{I},\tilde{K}_\mathrm{II}) + L_{t_2}(\pmb{w},\pmb{b},\tilde{K}_\mathrm{I},\tilde{K}_\mathrm{II}),$$

where $L_{PDE}^{(1)}$ and $L_{PDE}^{(2)}$ denote the loss terms of the two PDEs, $L_{u_1}$ and $L_{u_2}$ are loss terms correspond to displacement boundary conditions, $L_{t_1}$ and $L_{t_2}$ are loss terms that penalize the traction boundary conditions. The loss function measures how well the enriched solutions $u_1(\pmb{x},\pmb{w},\pmb{b},\tilde{K}_\mathrm{I},\tilde{K}_\mathrm{II})$ and $u_2(\pmb{x},\pmb{w},\pmb{b},\tilde{K}_\mathrm{I},\tilde{K}_\mathrm{II})$ satisfy the underlying PDEs and the corresponding boundary conditions. The parameters $\pmb{w}$, $\pmb{b}$, $\tilde{K}_\mathrm{I}$ and $\tilde{K}_\mathrm{II}$ can be trained by minimizing the loss function via various gradient-based optimization techniques. In our computations, we train the network using the MATLAB function "fmincon" which is designed to find the minimum of a constrained nonlinear/linear multi-variable functions. To implement PINNs efficiently, it is convenient to use various available algorithms based on the current machine learning (ML) libraries. In the end of this paper, we provide a self-contained MATLAB code to show the interested reader how to train a PINNs to numerically compute the crack problems by using the MATLAB. We restrict the description of the implementation to in-plane cracks in two dimensions. The above enrichments concepts can be easily extended to planar cracks in three dimensions but the implementation is more difficult.

For a fractured body, the displacement fields are discontinuous across the crack faces. Accordingly, special treatments are required to modify such discontinuous property. One direct way is to use the multi-domain technique for removing such discontinuity [14, 25]. The cracked material should be divided into two subdomains along the crack faces and the enriched PINNs model presented in this paper is then applied separately to each subdomain. The traction-free conditions are specified on both of the upper and lower crack-faces, where the displacement boundary conditions are discontinuous. On the bounded parts of the interface, the traction equilibrium and displacement continuity conditions should be satisfied. The finally displacement fields can be numerically solved by assembling equations written for each subdomain, based on the equilibrium and continuous conditions along the interface of the cracked material.

The modes I and II SIFs can be finally calculated by evaluating the displacement or stress fields at the crack-tip region, and by substituting these near-tip variables into relations -. For example, one direct way of computing the SIFs is to use the crack-opening-displacements (CODs) at the nodal points fallen behind the crack-tip, which results in:

$$K_I^* = \frac{\mu}{\kappa+1}\sqrt{\frac{2\pi}{r}}\left[(u_2)_{\theta=\pi} - (u_2)_{\theta=-\pi}\right], \tag{30}$$

$$K_{\text{II}}^* = \frac{\mu}{\kappa+1}\sqrt{\frac{2\pi}{r}}\left[(u_1)_{\theta=\pi} - (u_1)_{\theta=-\pi}\right], \tag{31}$$

where $K_{\text{I}}^*$ and $K_{\text{II}}^*$ are the estimated SIF results at nodal point with distance $r$ from the crack-tip. Since the SIFs correspond to the case of $r \to 0$, the final SIF results can be calculated by using the well-known displacement extrapolation method (DEM) [26, 27]. The main idea of the DEM is that the slope of the $K_{\text{I}}^*$ or $K_{\text{II}}^*$ curve, as function of the distance function $r$ for a fixed value of $\theta$, will rapidly approach to a constant with the value of $r$ increases. Therefore, the intercept of the tangent of the $K_{\text{I}}^*$- or $K_{\text{II}}^*$-curve with the $x_2$-axis (for the case of $r \to 0$) can be used as the final results of the SIFs (see the following Section).

## 5. Numerical examples and discussions

Four benchmark examples are analyzed numerically in this section to demonstrate the performance of the PINNs together with the crack-tip enrichment functions. The first two examples are chosen to illustrate the accuracy of the present algorithm for pure modes I and II fracture analysis of a finite plate with a center-crack, where the numerical simulation is compared with the corresponding theoretical solution. The effects of the network architectures, such as the number of hidden layers, the number of neuron units in each layer and the choice of activation function $\sigma(\cdot)$, on the overall accuracy of the PINNs models are carefully investigated. The last two examples are chosen to demonstrate the robustness of the enriched PINNs model in simulation of fracture problems with an edge and inclined cracks. The numerical simulation is compared with the boundary element method by evaluation of the SIFs for different crack angles. In all benchmarks considered in this section, the total number of training data is relatively small (up to a few thousand points). All numerical experiments are carried out by using MATLAB R2021b, running on a computer with i7 2.90GHz CPU, 32GB memory, 500 GB hard drive and a Windows11 operating system. The resulting relative error is calculated by $|K_{num} - K_{exact}|/|K_{exact}|$, where $K_{num}$ and $K_{exact}$ denote the numerical and exact stress solutions, respectively.

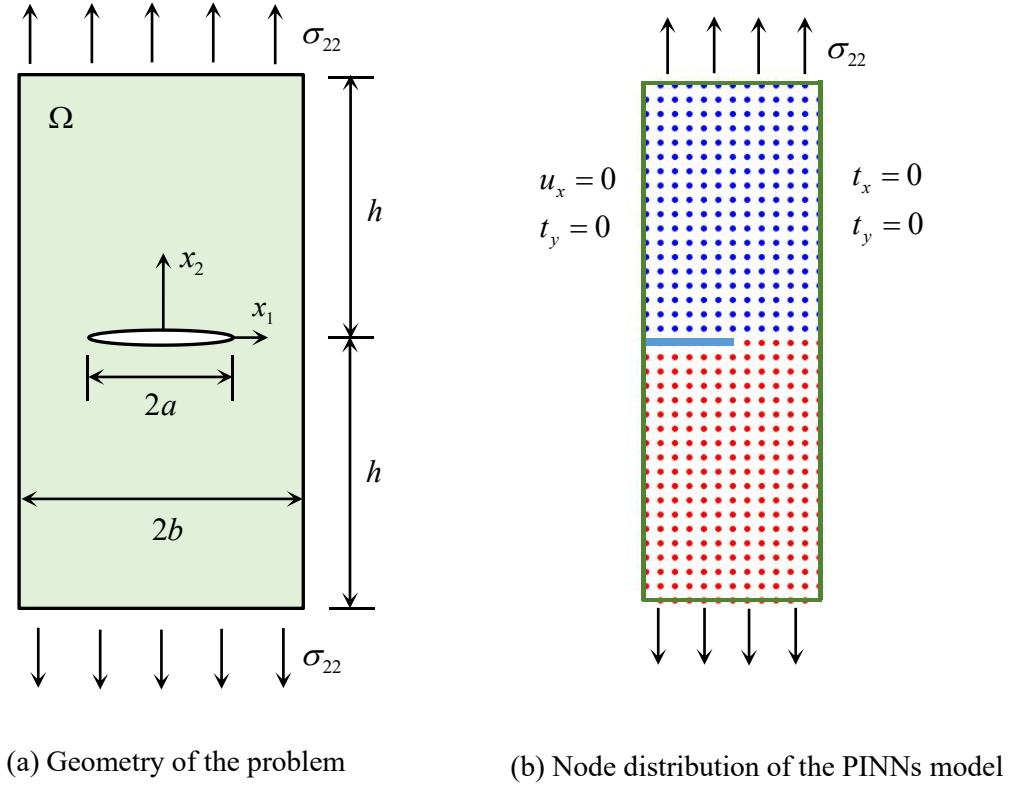

(a) Geometry of the problem

(b) Node distribution of the PINNs model

**Fig. 4.** A center-cracked plate with remote tensile stress loading. (a) Geometry of the problem, and (b) node distribution of the PINNs model.

*5.1. A center-cracked plate with remote tensile stress loading*

The first example is chosen to illustrate the performance of the proposed PINNs formulation for modeling a pure mode I fracture problem. The geometry, boundary conditions and the node distribution of the PINNs model are shown in Fig. 4. The plate is in plane strain condition with Poisson's ratio $v = 0.3$ and the plate subjected to the prescribed tensile stress loading $\sigma_{22} = 1$ at the top and bottom edges. The crack has a half-length $a = 1$ which is located in the center of the plate. The width and height of the plate are $2b$ and $2h$, respectively, where we take $b/a = 2$ and $h/a = 6$ for our particular example. The effect of $h$ is practically negligible for $h/b \geq 1$.

Since the problem has symmetry about the $x_2$- axis, the computational domain shown in Fig. 4(b) is considered. The theoretical solution of the SIF $K_\mathrm{I}$ for this problem is [20]:

$$K_\mathrm{I} = \sigma_{22}\sqrt{\pi a}\left[1 - 0.025\left(\frac{a}{b}\right)^2 + 0.06\left(\frac{a}{b}\right)^4\right]\sqrt{\sec\frac{\pi a}{2b}}. \quad (32)$$

Training the model using the PINNs requires a set of collocation points in order to enforce the PDEs and the corresponding boundary constraints. The easiest way for this example is to choose all points on an $m \times n$ grid which fall within the domain. Here, the network is trained by using a $30 \times 180$ rectangle grid, without using additional nodal refinement in the crack-tip region.

**Table 1.** Performance of the present PINNs model using different network architectures (mode-I).

| Network architecture | | Exact solution | Enriched PINNs | Relative errors |
|---|---|---|---|---|
| 3 hidden layers | 10 neurons | | 2.0306 | $3.4213 \times 10^{-2}$ |
| | 15 neurons | 2.1025 | 2.0444 | $2.7632 \times 10^{-2}$ |
| | 20 neurons | | 2.0722 | $1.4440 \times 10^{-2}$ |
| 4 hidden layers | 10 neurons | | 2.0502 | $2.4894 \times 10^{-2}$ |
| | 15 neurons | 2.1025 | 2.0853 | $8.1850 \times 10^{-3}$ |
| | 20 neurons | | 2.0978 | $2.2702 \times 10^{-3}$ |
| 5 hidden layers | 10 neurons | | 2.0732 | $1.3972 \times 10^{-2}$ |
| | 15 neurons | 2.1025 | 2.0889 | $6.4857 \times 10^{-3}$ |
| | 20 neurons | | 2.0959 | $3.1542 \times 10^{-3}$ |
| 6 hidden layers | 10 neurons | | 2.0765 | $1.2405 \times 10^{-2}$ |
| | 15 neurons | 2.1025 | 2.0936 | $4.2652 \times 10^{-3}$ |
| | 20 neurons | | 2.0996 | $1.4178 \times 10^{-3}$ |

The network architecture of a PINNs model can be crucial to its success. Frequently, different applications require different network architectures. Here, the performance of the present PINNs model is evaluated for different choices of the underlying network architecture, obtained by varying the number of hidden layers as well as neuron units per layer. Table 1 summarizes the corresponding relative errors for the predicted SIF results, where all computations are performed with 2500 iteration steps and by using a $30 \times 180$ rectangle grid. The activation function used for each hidden layer is taken to be $\sigma(z) = z/(1+e^{-z})$ (Swish function). It can be observed from Table 1 that the predicted SIF results calculated by using the present PINNs model agree very well with the reference results. We can also observe that the present PINNs model appears to be very robust and stable with respect to the underlying network architectures and shows a consistent trend in improving the prediction accuracy as the number of hidden layers and neurons is increased. The results clearly demonstrate the accuracy and validity of the present PINNs for in-plane crack analysis. It is also noted that, with the enrichment, the PINNs model is able to capture the near-tip singularity and eliminate oscillations without using extra nodal refinement in the crack-tip region. *A self-contained MATLAB code accompanying this example can be found in the end of the paper, in order to provide sufficient information for the reconstruction of the results by interested readers.*

**Table 2.** Performance of the PINNs model using different activation functions.

| | Sigmoid | Tanh | Swish | Arctan | Softplus |
|---|---|---|---|---|---|
| $\sigma(z)$ | $\dfrac{1}{1+e^{-z}}$ 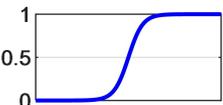 | $\dfrac{e^z - e^{-z}}{e^z + e^{-z}}$ 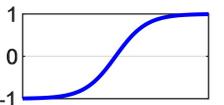 | $\dfrac{z}{1+e^{-z}}$ 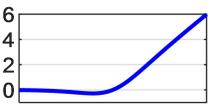 | $\arctan(z)$ 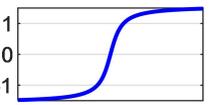 | $\ln(1+e^z)$ 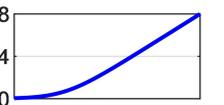 |
| Relative errors | $6.6414 \times 10^{-3}$ | $5.7186 \times 10^{-3}$ | $6.4857 \times 10^{-3}$ | $5.1468 \times 10^{-3}$ | $8.1370 \times 10^{-3}$ |

Next, we fix the network architecture to 5 fully-connected hidden layers with 15 neurons per layer and learn the network using different activation functions. Table 2 shows the relative errors of the predicted SIF by using different activation functions after 2500 iteration steps. It can be observed that the performance of the present PINNs model is relatively insensitive to the choice of the activation functions where all cases are able to correctly identify the SIF solutions with very good accuracy, yielding the relative errors ranging between $5\times 10^{-3}$ and $7\times 10^{-3}$.

As described in Section 4, the final SIF results can be calculated by using the displacement extrapolation method, that is, by extrapolating the straight portion of the $K_I^*$-curve back to the $x_2$-axis. The plot of the $K_I^*$-curve, calculated by using the crack-opening-displacements at the nodal points fallen behind the crack-tip, can be found in Fig. 5. It can be observed that the slope of the $K_I^*$-curve rapidly approaches a constant with the value of $r$ increases. The intercept of the tangent of the curve with the $x_2$-axis can be used as the final solution of the SIF $K_I$.

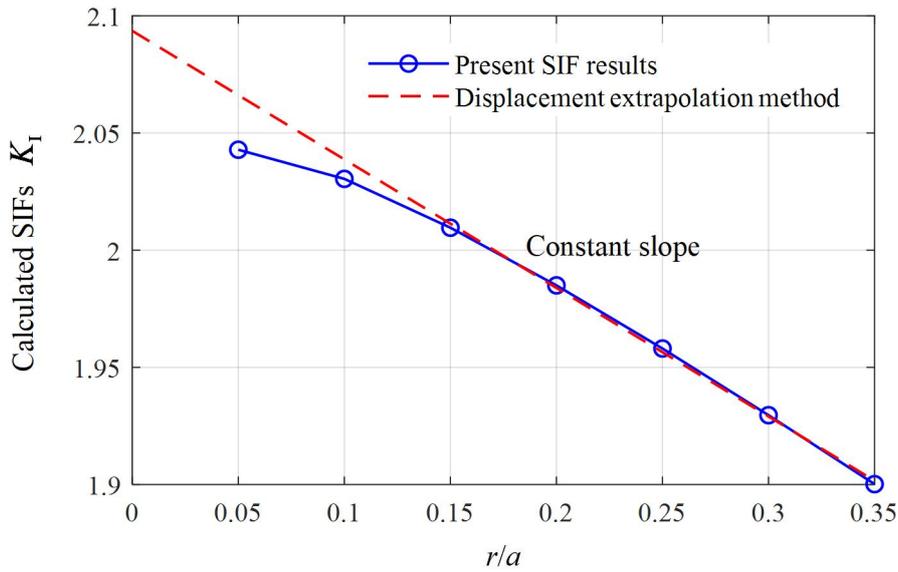

**Fig. 5.** The estimated results of the mode-I SIF $K_I$ calculated by using the displacement extrapolation method.

In Fig. 6, we present the variation of the estimated SIF $K_I$ with respect to various values of crack-length ratios $b/a$, where the value of $a$ is fixed as $a=1$ while the ratio $b/a$ reduces as $b$ decreases. The network is trained here by using 5 fully-connected hidden layers with 20 neurons per layer and 2500 iteration steps. We can observe that the predicted SIF results calculated by using the present PINNs model are excellently consistent with the corresponding reference results for a wide range of crack-length ratios. Again, it is noted that the total number of collocation points used to train the model is kept the same across the entire range of $b/a$.

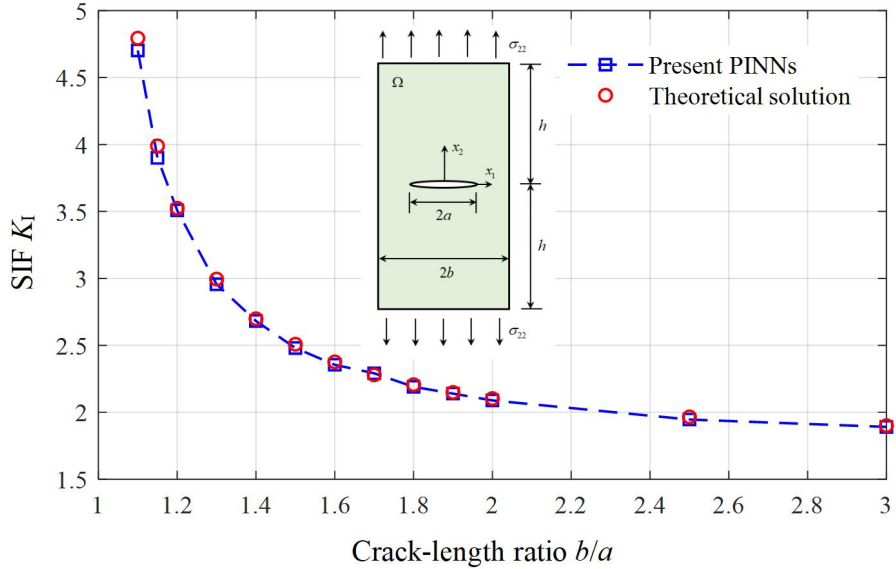

**Fig. 6.** Variation of the estimated SIF $K_I$ for various values of relative crack-length ratio $b/a$.

*5.2. A center-cracked plate with remote shear stress loading*

The second example is chosen to illustrate the performance of the present PINNs for modeling a pure mode II fracture problem. The geometry, boundary conditions and the node distribution of the PINNs model are shown in Fig. 7. The plate is subjected to a remote shear stress loading $\tau_{xy} = 1$. The half crack-length is $a = 1$ and the half width and height of the plate are taken to be $b = 2$ and $h = 6$, respectively. Again, the plate is in plane strain condition with Poisson's ratio $v = 0.3$. Because of the symmetry, the computational domain shown in Fig. 7 is considered. Here, the network is trained by using a $40 \times 200$ rectangle grid, without using additional nodal refinement in the crack-tip region. The exact solution of the mode-II SIF $K_{II}$ can be found in Ref. [20] as follows:

$$K_{II} = \tau_{xy}\sqrt{\pi a}\left[1 - 0.025\left(\frac{a}{b}\right)^2 + 0.06\left(\frac{a}{b}\right)^4\right]\sqrt{\sec\frac{\pi a}{2b}}. \tag{33}$$

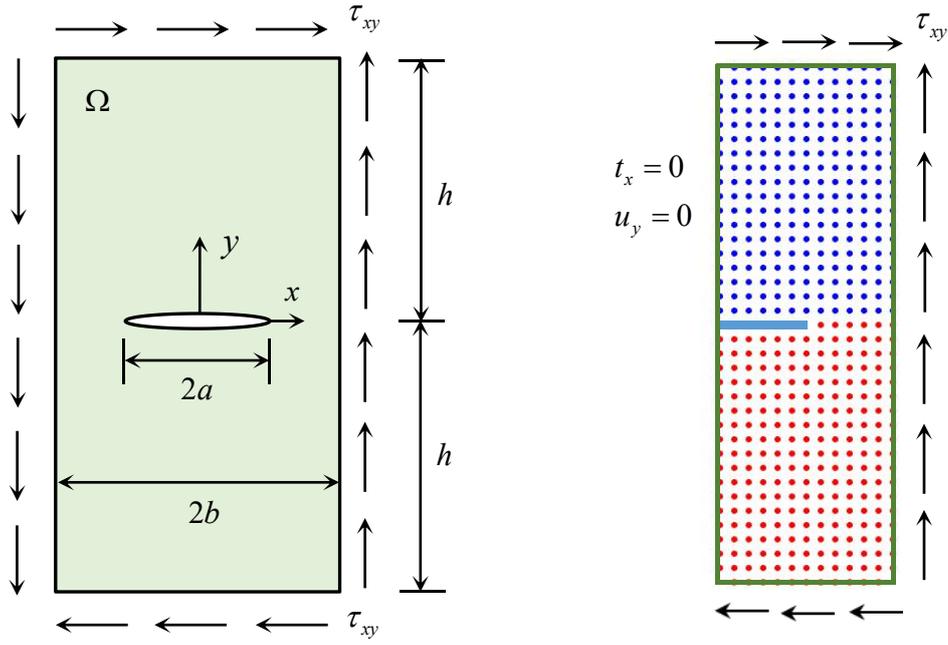

**Fig. 7.** A center-cracked plate subjected to remote shear stress loading.

**Table 3.** Performance of the present PINNs model using different network architectures (mode-II).

| Network architecture | | Exact solution | Enriched PINNs | Relative errors |
|---|---|---|---|---|
| 3 hidden layers | 10 neurons |  | 2.0748 | $1.3179 \times 10^{-2}$ |
|  | 15 neurons | 2.1025 | 2.1106 | $3.8451 \times 10^{-3}$ |
|  | 20 neurons |  | 2.0960 | $3.1075 \times 10^{-3}$ |
| 4 hidden layers | 10 neurons |  | 2.0833 | $9.1584 \times 10^{-3}$ |
|  | 15 neurons | 2.1025 | 2.1084 | $2.8004 \times 10^{-3}$ |
|  | 20 neurons |  | 2.0970 | $2.6342 \times 10^{-3}$ |
| 5 hidden layers | 10 neurons |  | 2.1170 | $6.8960 \times 10^{-3}$ |
|  | 15 neurons | 2.1025 | 2.1111 | $4.0788 \times 10^{-3}$ |
|  | 20 neurons |  | 2.0989 | $1.7530 \times 10^{-3}$ |
| 6 hidden layers | 10 neurons |  | 2.1157 | $6.2659 \times 10^{-3}$ |
|  | 15 neurons | 2.1025 | 2.1081 | $2.6421 \times 10^{-3}$ |
|  | 20 neurons |  | 2.1058 | $1.5367 \times 10^{-3}$ |

Table 3 summarizes the relative errors of the predicted SIFs by using different number of hidden layers and different number of neuron units per layer. All computations are performed with 2500 iteration steps and by using a $40 \times 200$ rectangle grid. The activation function used here for each hidden layer is taken to be $\sigma(z) = z/(1+e^{-z})$ (Swish function). It can be observed from Table 3 that the predicted SIF results calculated by using the present PINNs model agree very well with the reference results. Again, we can also observe that the present PINNs model shows a consistent trend in improving the prediction accuracy as the number of hidden layers and neurons is increased. The plot of the $K_{II}^*$-curve, calculated by using the crack-opening-displacements at the nodal points fallen behind the crack-tip, can be found in Fig. 8. It can be observed that the slope of the $K_{II}^*$-curve rapidly approaches a constant with the value of $r$ increases. The intercept of the tangent of the curve with the $x_2$-axis can be used as the final solution of the SIF $K_{II}$. In Fig. 9, we present the variation of the estimated SIF $K_{II}$ with respect to various values of crack-length ratios $b/a$. The network is trained here by using 5 fully-connected hidden layers with 20 neurons per layer and 2500 iteration steps. We can observe that the SIF results calculated by using the present PINNs model are excellently consistent with the corresponding exact results.

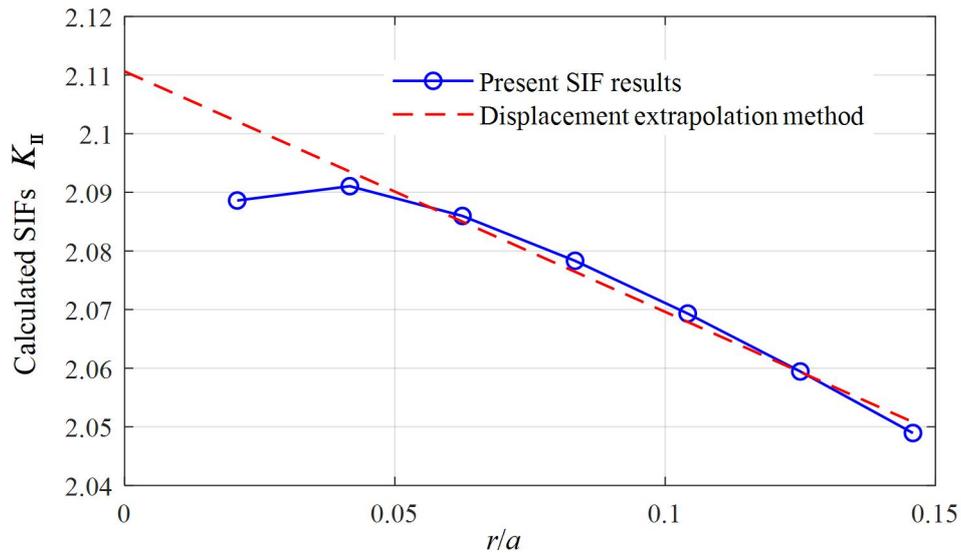

**Fig. 8.** The estimated results of the mode-II SIF $K_{II}$ calculated by using the displacement extrapolation method.

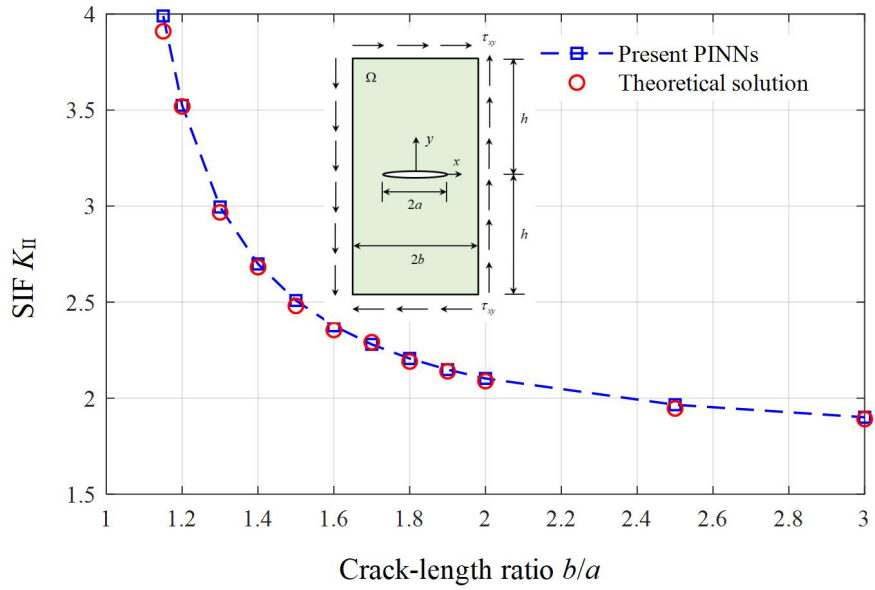

**Fig. 9.** Variation of the estimated SIF $K_{II}$ for various values of relative crack-length ratio $b/a$.

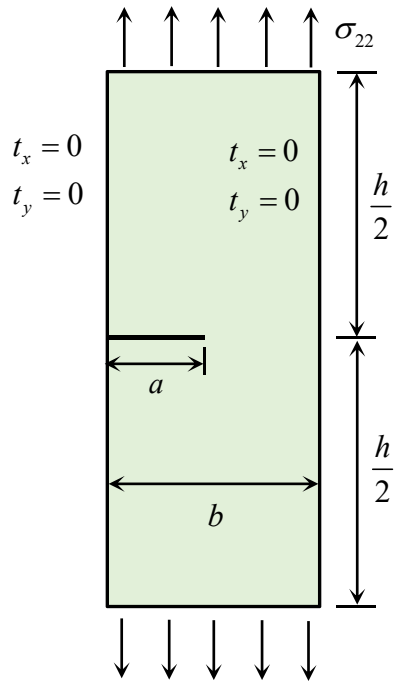

**Fig. 10.** An edge-cracked plate.

## *5.3. An edge-cracked plate subjected to remote tensile stress loading*

As shown in Fig. 10, the third problem considered is an edge-cracked plate subjected to uniform tensile stress loading $\sigma_{22}$. The length of the crack is $a$ and the principal dimensions of the plate are $b=1$ in width and $h=2b$ in length. Again, this example is assumed to be under plane strain condition with Poisson's ratio $v=0.3$. An empirical formula for the crack opening displacement at the edge of the plate for any $a/b$ is [20]:

$$\left(\Delta u_2\right)_{edge} = \left(u_2\right)_{\substack{\theta=\pi \\ r=a}} - \left(u_2\right)_{\substack{\theta=-\pi \\ r=a}} = \frac{4\sigma_{22}a\left(1-v^2\right)}{E}\left[\frac{1.46+3.42\left(1-\cos\left(\pi a/2b\right)\right)}{\left(\cos\left(\pi a/2b\right)\right)^2}\right], \quad (34)$$

where $E$ denotes the Young's modulus. This formula can be used to check the validity of the PINNs solutions since there are no analytical solutions available in the literature, to the best of the authors' knowledge, for this benchmark example. In the following, we train the network by using 6 fully-connected hidden layers with 15 neurons per layer. The activation function used for each hidden layer is taken to be $\sigma(z)=z/(1+e^{-z})$. All computations are performed with 2500 iteration steps and by using a $20\times100$ rectangle grid.

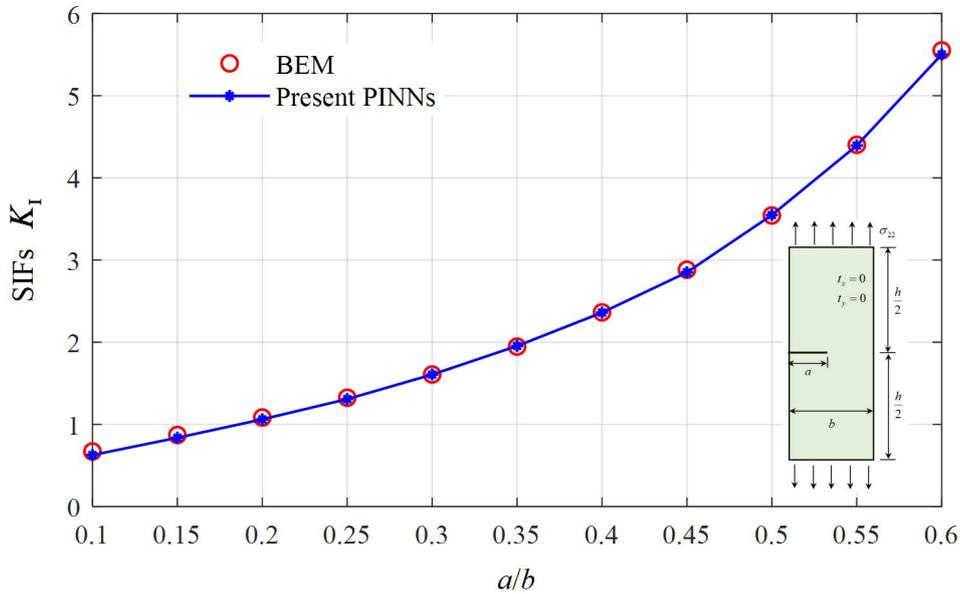

Fig. 11. Comparison between the BEM and the PINNs results for predicted SIFs $K_I$.

The variation of the estimated SIFs $K_I$ is plotted in Fig. 11 as the relative crack length $a/b$ changes from 0.1 to 0.6. This benchmark was also solved with the BEM [28] which has been widely acknowledged as one of the most effective approaches to the fracture mechanics analysis. The BEM results were obtained here based on the use of the multi-domain method together with a special crack-tip element technique, see Ref. [28]. As can be seen from Fig. 11, the numerical solutions calculated by using the present PINNs model are excellently consistent with the corresponding BEM solutions in a wide range of $a/b$. The results clearly demonstrate the validity and accuracy of the present PINNs for in-plane crack analysis. It is noted again that the present PINNs model is able to capture the near-tip singularity and eliminate oscillations without using extra nodal refinement in the crack-tip region. The computing time and the memory requirements are therefore quite modest. We can also observe from Fig. 11 that the SIF $K_\mathrm{I}$ increases monotonically with the increase in the crack length ratio $a/b$. Table 4 presents the prediction results for the normalized crack opening displacement $\left(\Delta u_2\right)_{edge} E / \sigma_{22} a \left(1-v^2\right)$ at the edge of the plate in comparison with the reference results calculated by using the empirical formula (34). We can observe that the present results show excellent agreements with the corresponding reference solutions, which verifies again the accuracy of the developed algorithm for in-plane crack analysis.

**Table 4.** Comparison of the normalized crack opening displacement at the edge of the plate with reference results calculated by using the empirical formula (34).

| $a/b$ | 0.1 | 0.2 | 0.3 | 0.4 | 0.5 | 0.6 |
|---|---|---|---|---|---|---|
| Reference results [20] | 6.1591 | 7.1967 | 9.2342 | 12.9144 | 19.6935 | 33.2254 |
| Present PINNs | 6.1098 | 7.2815 | 9.3029 | 12.8423 | 19.5503 | 33.1052 |

The variation of the normalized crack opening displacement $\left(\Delta u_2\right) E / \sigma_{22} a \left(1-v^2\right)$ is illustrated in Fig. 12, with $a=0.5$ and $b=1$. The "exact" solution of the normalized crack opening at the edge of the plate can be calculated by using the empirical formula (34), i.e., $(\Delta u_2)_{edge} E / \sigma_{22} a (1-v^2) = 19.69$. We can observe from Fig. 12 that the normalized crack opening displacement approaches 19.69 as $x_1 \to 0$ (the left-edge of the plate). This can be used to check the validity of the present PINNs solutions. Again, the plot of the $K_I^*$-curve calculated at nodal points near the crack-tip region can be found in Fig. 13, with $a = 0.5$ and $b = 1$.

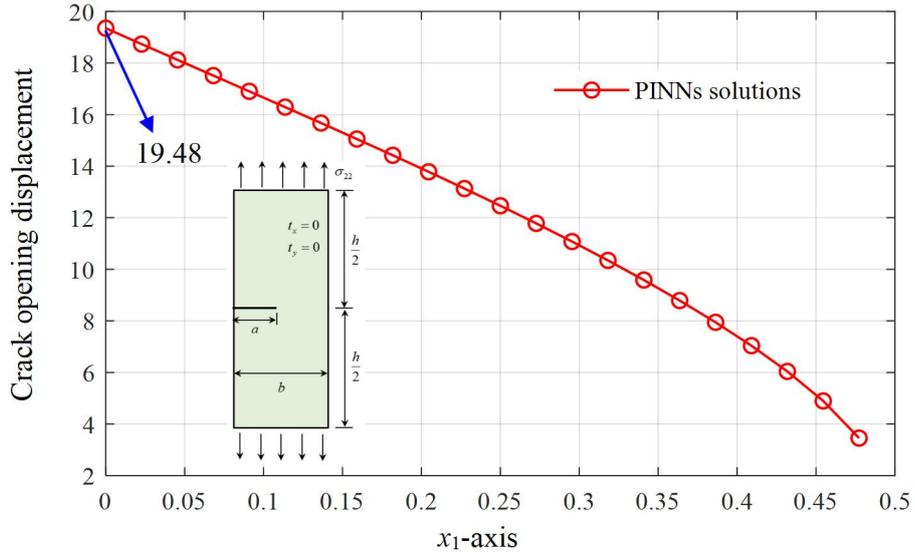

**Fig. 12.** Variation of the normalized crack opening displacement with $a = 0.5$ and $b = 1$.

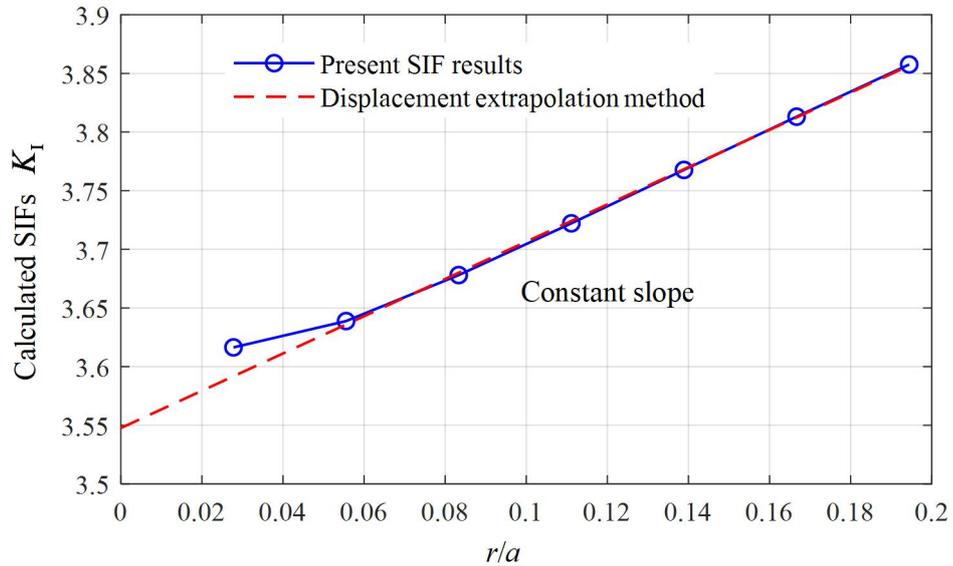

**Fig. 13.** The estimated results of the mode-I SIF $K_I$ calculated by using the displacement extrapolation method ($a = 0.5$ and $b = 1$).

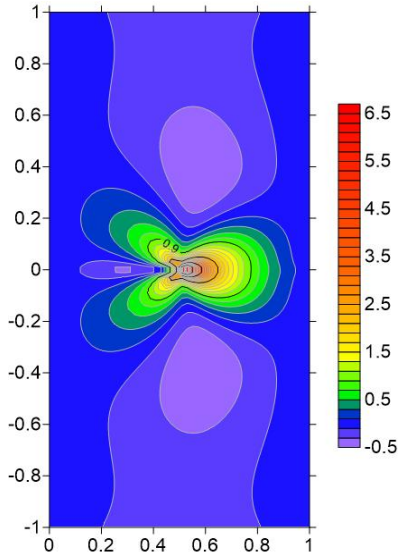 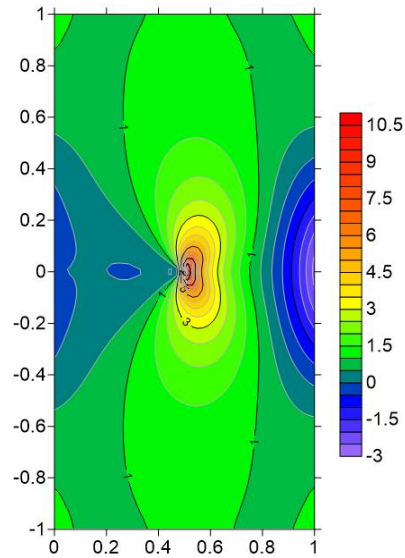 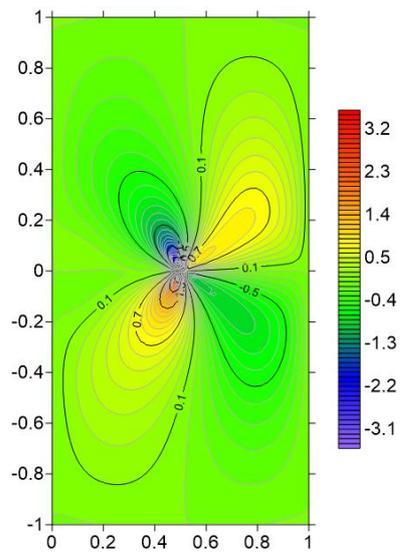

**Fig. 14.** Colored contour plot for calculated stresses $\sigma_{11}$.

**Fig. 15.** Colored contour plot for calculated stresses $\sigma_{22}$.

**Fig. 16.** Colored contour plot for calculated stresses $\sigma_{12}$.

The colored contour plots for calculated stresses $\sigma_{11}$, $\sigma_{22}$ and $\sigma_{12}$ are plotted in Figs. 14-16, respectively, in the case of $a = 0.5$ and $b = 1$. The stress singularities in the near-tip regions can be observed from these figures.

### *5.4. A finite plate with a slant crack*

The final example is chosen to illustrate the performance of the proposed PINNs formulation for modeling a mixed mode fracture problem, i.e., the calculation of SIFs $K_\mathrm{I}$ and $K_\mathrm{II}$ of a finite plate with a slant crack [11, 29]. As shown in Fig. 17, the half-length of the crack, the width and length of the plate are $a = 1$, $H = 4$ and $2H$, respectively. This benchmark is assumed to be under plane stress condition with the Poisson's ratio $v = 0.3$. The numerical results are carried out for different values of the alant angle $\theta$ of the crack-faces. Here, the network is trained by using a $40 \times 200$ rectangle grid, and by employing 6 fully-connected hidden layers with 20 neurons per layer and 2500 iteration steps.

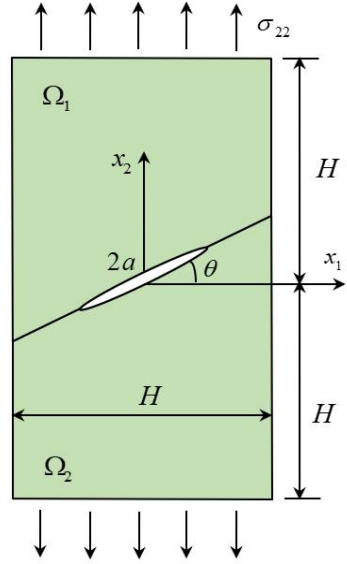

**Fig. 17.** A finite plate with a slant crack.

The results of the SIFs $K_\mathrm{I}$ and $K_\mathrm{II}$ obtained for various values of $\theta$ are listed in Table 5. The numerical results carried out by using the BEM [11] are also given for the purpose of comparison. Again, as expected, the present PINNs model gives very accurate results in excellent agreement with the solutions calculated by using the BEM. This illustrates again that the present method can provide accurate and reliable solutions for in-plane crack analysis.

**Table 5.** Comparison of the SIFs $K_\mathrm{I}$ and $K_\mathrm{II}$ for different values of $\theta$.

| Methods | $K_\mathrm{I}$ | | | | $K_\mathrm{II}$ | | | |
|---|---|---|---|---|---|---|---|---|
| | Right-tip of the crack | | | | Right-tip of the crack | | | |
| | 15° | 30° | 45° | 60° | 15° | 30° | 45° | 60° |
| Present PINNs | 1.9715 | 1.6058 | 1.0904 | 0.5441 | 0.4696 | 0.8205 | 0.9618 | 0.8592 |
| BEM [11] | 1.9705 | 1.6020 | 1.0838 | 0.5494 | 0.4641 | 0.8180 | 0.9666 | 0.8551 |

## 6. Conclusions

In this paper, a new framework based on the physics-informed neural networks (PINNs) has been presented for modelling in-plane crack problems in the linear elastic fracture mechanics. The crack-tip asymptotic/enrichment functions were added to the standard PINNs approximation so that the enriched method is able to correctly solve crack problems with very high accuracy even using relatively small set of training data. The accuracy and efficiency of the present method are demonstrated by a class of representative benchmarks with different modes of loading types. Some attractive features of the present method compared with the existing techniques for fracture mechanics analysis can be summarized as follows:

(1) Instead of forming a mesh, the PINNs is meshless and can be trained on batches of randomly sampled collocation points.

(2) The PINNs model provides closed-form and differentiable solutions which are easily used in any subsequent calculation.

(3) The enriched PINNs presented in this paper can properly describe the local asymptotic properties of the near-tip displacement and stress fields without a high degree of nodal refinement. The method, therefore, enables a direct calculation of the SIFs without the usage of the classical techniques based on, for example, the path-independent contour integrals, the interaction integrals and the virtual crack extension technique.

(4) The method based on the PINNs can easily be used for dealing with crack problems with higher dimensions, since the number of training parameters can remain almost fixed as the problem dimensionality increases.

In addition to the above advantages, one of the main limitations of the PINNs-based methods is a lack of sufficient and rigorous mathematical analysis. Specifically, the performance of the method is evaluated using empirical observations without any rigorous mathematical proof. Although we have verified that such an approach can be applied to a wide range of applications, the convergence and stability of the method are worth further investigation. Although the present method is developed in the context of in-plane crack problems, its extension to other problems, for example, the bi-material interface crack problems, dynamic crack propagation and 3D crack problems, is fairly straightforward. The present work provides a solid basis for these interesting research topics. Results along these lines will be presented in the future.

## Acknowledgements

The work described in this paper was supported by the National Natural Science Foundation of China (Nos. 11872220, 12111530006), and the Natural Science Foundation of Shandong Province of China (Nos. 2019KJI009, ZR2021JQ02).

## Data Availability Statement

The data that support the findings of this study are available from the corresponding author upon reasonable request.

## Appendix A. The self-contained source code for benchmark 5.1 (MATLAB$^{®2021b}$)

The following codes show how to train physics informed neural networks to numerically compute the in-plane crack problems (example 5.1). The idea of part of the codes is from MathWorks at: https://ww2.mathworks.cn/help/deeplearning/ug/solve-partial-differential-equations-with-lbfgs-method-and-deep-learning.html.